\documentclass[letter,twocolumn]{jpsj2} 

\newcommand{\bcite}[1]{\cite{#1}}

\newcommand{\Ham}{{\cal H}}

\newcommand{\vect}[1]{{\mbox{\boldmath $#1$}}}

\newcommand{\figlabel}[1]{\label{fig:#1}}

\newcommand{\Eq}[1]{(\ref{eq:#1})}
\newcommand{\Fig}[1]{Fig.\ \ref{fig:#1}}
\newcommand{\Figure}[1]{Figure \ref{fig:#1}}

\newcommand{\deffig}[4]{
\begin{figure}[tb]
  \begin{center}
  \includegraphics[width=#3 \textwidth]{Fig/#2}
  \end{center}
  \caption{ \figlabel{#1} #4}
\end{figure}
}

\title{
  Quantum Critical Point of the $XY$ Model and
  Condensation of Field-Induced Quasiparticles in Dimer Compounds 
}
\author{
  Naoki \textsc{KAWASHIMA}
  \thanks{E-mail address: kawashima@issp.u-tokyo.ac.jp}
}
\inst{
  Department of Physics, Tokyo Metropolitan University, Hachioji, Tokyo 192-0357, Japan; \\
  Institute for Solid State Physics, University of Tokyo, Kashiwa, Chiba 277-8581, Japan
}
\abst{
The quantum critical point of the three-dimensional $XY$ model 
in a symmetry-preserving field is investigated.
The results of Monte Carlo simulations with the directed-loop algorithm 
show that the quantum critical behavior is characterized
by the mean-field values of critical exponents.
The system-size dependence of various quantities is compared to
a simple field-theoretical argument that supports
the mean-field scaling.
}
\kword{quantum spin system, Heisenberg model, quantum Monte Carlo,
XY model, XXZ model, cluster algorithm, loop algorithm,
worm algorithm, directed-loop algorithm}

\begin{document}
\maketitle


Recent developments of the high-field magnets and related 
experimental techniques make it possible to explore 
the magnetic phenomena revealed only by a strong magnetic field.
A typical example is the dimer compounds such as 
TlCuCl$_3$\bcite{TlCuCl1999,TlCuCl2001}, KCuCl$_3$\bcite{KCuCl} and
BaCuSi$_2$O$_6$\bcite{JaimeETAL2004}.
These compounds consist of dimers each being an
antiferromagnetically coupled pair of spins.
Without a strong magnetic field, spins form singlet pairs and 
the compounds are magnetically inactive.
It was suggested\bcite{TachikiY1970,TachikiY1970b} that 
when a strong magnetic field is applied to such a dimer system,
triplet excitations are induced and the system
may exhibit a phase transition at a finite temperature.
This phase transition can be most naturally interpreted
\bcite{NikuniOOT2000,Rice2002} as a condensation transition of 
the excitations that behave as bosonic quasiparticles.
Indeed, the results of the Hartree-Fock (HF) calculation
\bcite{NikuniOOT2000} of the diluted Bose gas
qualitatively agree with the experimental observations,
such as the characteristic temperature dependence of the 
magnetization and the algebraic temperature dependence of the 
critical magnetic field.
Results of another recent experiment\bcite{RueggETAL2003}
also identified the phase transition as the Bose-Einstein condensation (BEC).
Even a quantitative agreement was obtained\bcite{JaimeETAL2004}
recently for BaCuSi$_2$O$_6$ for which the
temperature dependence of the experimentally measured specific heat 
agreed with the Monte Carlo simulation results of the effective 
hard-core boson model, which showed a clear `lambda' peak.
Thus, the nature of the transition as a condensation transition
of the triplet excitations has been established beyond reasonable doubt.

However, an unsettling disagreement has been left unsolved,
in estimates of the critical exponent that characterizes
the temperature dependence of the critical magnetic field.
The experiments and the theories both suggest the algebraic dependence,
$
  H_c(T)-H_c(0) \propto T^{\phi}.
$
However, the experimental estimates of $\phi$ range from 
$1.7$ \cite{TlCuCl1999} to $2.0$,\cite{TlCuCl2001}
whereas the theoretical estimate (namely, the HF value) is
$\phi = 1.5$.\cite{NikuniOOT2000}
A Monte Carlo simulation\cite{NohadaniWNH2003} was performed recently
on the effective spin model for TlCuCl$_3$.
A temperature dependent effective exponent $\phi(T)$
was defined such that it characterizes the $H_c-T$ curve in a certain
finite temperature-range centered at $T$.
The simulation result showed that $\phi(T)$ is greater than 1.5
but seems to approach 1.5 as the temperature is decreased.
Moreover, the HF calculation was elaborated 
recently\bcite{MisguichO2004} based on a more realistic dispersion 
relation for bosons determined by experiments.
While it is doomed to yield only mean-field values of critical exponents,
which obviously differ from the correct ones for a finite-temperature BEC transition,
it still provides a fairly good approximation for the value of the critical temperature.
Therefore, these recent simulation and theoretical studies may be suggesting that
the failure of the HF approximation in the finite-temperature critical phenomena
does not necessarily mean the failure in the quantum critical phenomena.
In particular, the mean-field value $\phi = 1.5$ may be the correct 
and exact value near the zero temperature.
This statement, however, has not been made conclusive 
because of the uncontrollability of the approximation and/or
the various limitations (such as the system size and the
temperature range) in the Monte Carlo simulation.

In this letter, we present a simple theoretical argument,
from which one can conclude that the mean-field values of the
critical exponents are exact for the quantum critical point (QCP),
i.e., $T=0$, $H=H_c(0)$,
and therefore the inaccuracy of the HF approximation is not
the reason for the disagreement.
To demonstrate the validity of the argument,
we also show some results of a quantum Monte Carlo
simulation on the three-dimensional $XY$ model with
a magnetic field.


We start with the model in which each spin (e.g., each Cu spin in 
TlCuCl$_3$) is treated explicitly:
$$
  \Ham =     \sum_i \hat J \vect{s}_{i1}\cdot\vect{s}_{i2}
           + \sum_{(ij)}\sum_{\alpha,\beta} \hat J'_{i\alpha,j\beta} 
             \vect{s}_{i\alpha}\cdot\vect{s}_{j\beta}
           - \sum_{i\alpha} \hat H s^z_{i\alpha}.
$$
Here, $\hat J>0$ is the antiferromagnetic coupling constant that binds
a pair of spins, $i$ the index of the dimer and 
$\vect{s}_{i1}$ and $\vect{s}_{i2}$ the two spins that form
the pair $i$.
The lattice structure and the magnitude of the
coupling constants depend on the compound.
In the case of TlCuCl$_3$\cite{TlCuCl1999}, for example,
the lattice is composed of double chains of spins (i.e., two-leg ladders) 
with relatively weak interladder couplings.
The dominant couplings are theose corresponding to
rungs of the ladder, which are represented by $\hat J$ above.
Below, we assume that the interdimer couplings
$\hat J'_{i\alpha,j\beta}$ 
are much smaller (in magnitude) than $\hat J$, 
and that the lattice is three-dimensional.
Real materials, such as TlCuCl$_3$, are often strongly anisotropic.
The spatial anisotropy is, however, irrelevant in the present letter
since we focus on the critical properties,
in particular, the critical properties near $T=0$,
where the three-dimensionality manifests itself.

Following Tachiki and Yamada\cite{TachikiY1970,TachikiY1970b}, we neglect the two of the
triplet states that are not favored by the magnetic field.
The remaining two states, the triplet state that is
favored by the magnetic field and the singlet state,
are regarded as the up and down states, respectively, of 
an effective $S=1/2$ spin.
The resulting Hamiltonian in terms of the effective spin 
$\vect{S}_i$ is\cite{TachikiY1970,TachikiY1970b}
\begin{equation}
  \Ham =
           \sum_{(ij)} \tilde J_{ij}
           \left( S_i^x S_j^x + S_i^y S_j^y + \frac12 S_i^z S_j^z \right)
         - \sum_i \tilde H_i S_i^z,
  \label{eq:ReducedHamiltonian}
\end{equation}
where $\tilde J_{ij}$ and $\tilde H_i$ can be expressed as simple
linear combinations of $\hat J$, $\hat H$ and $\hat J'_{i\alpha,j\beta}$.
We here assume that the effective Hamiltonian
is translationally invariant and nonfrustrated.
The experimental estimations of the lattice structure and
the coupling constants suggest 
that these assumptions are valid, at least approximately,
for several compounds such as TlCuCl$_3$\bcite{TlCuCl2002}
and BaCuSi$_2$O$_6$\cite{BaCuSiO}.

Thus, we have arrived at the uniform $S=1/2$ $XXZ$ model 
in three dimensions with a spatial anisotropy 
and an easy-plane spin anisotropy.
The common belief is that the spatial anisotropy 
does not affect the critical properties of the model
both at the finite temperature transition and the QCP.
In addition, the $S^z$-$S^z$ term can be dropped without
affecting the critical properties, since it does not change
the type of spin anisotropy.
Therefore, we can further simplify the model
without changing the critical properties:
\begin{equation}
  \Ham =
         - J \sum_{(ij)}
             \left( S_i^x S_j^x + S_i^y S_j^y  \right)
         - H \sum_i S_i^z,
  \label{eq:XYHamiltonian}
\end{equation}
i.e., the $XY$ model Hamiltonian.
The sign of the coupling constant $J$ can be switched by changing
the spin representation bases. Therefore, we assume $J>0$ and $H>0$ 
without loss of generality in what follows.

If we consider the effective Hamiltonian as the sum of pair 
Hamiltonians $H_{ij}$, 
it is straightforward to show that the state in which the two 
spins $S_i$ and $S_j$ are aligned with the magnetic field is
the unique ground state of $H_{ij}$ when $H > dJ$.
Therefore, the state with all spins pointed up
is the unique ground state of the entire system for $H > dJ$.
On the other hand, it can also be shown easily that 
mixing with one-magnon states decreases the energy of 
the all-up state when $H < dJ$.
Therefore, for the present system, the quantum critical point 
is located exactly at $H=dJ$.

While the Hamiltonian \Eq{XYHamiltonian} is the one that we simulate with the
quantum Monte Carlo method, we can further reform it by using
the bosonic representation
$S_i^+ = b^{\dagger}_i$, $S_i^- = b_i$, and $S_i^z = \hat n_i - \frac12$
with the constraint
\begin{equation}
  \hat n_i \equiv b^{\dagger}_i b_i = 0, 1.
  \label{eq:Condition}
\end{equation}
Then we obtain
\begin{equation}
  \Ham =
         - t \sum_{(ij)}
             \left( b_i^{\dagger} b_j + b_j^{\dagger} b_i  \right)
         - \mu \sum_i \hat n_i
         + \Lambda \sum_i \hat n_i (\hat n_i - 1),
  \label{eq:BosonicHamiltonian}
\end{equation}
with new parameters $t$ and $\mu$ corresponding to the old ones 
as $t \sim J/2$ and $\mu \sim H$.
The $\Lambda$-term imposes 
the condition \Eq{Condition}, approximately.

Generally, the replacement of the original
`hard' constraint by a `soft' constraint
may affect the critical properties around the QCP.
In the present case, however, we consider that this modification 
does not change the essential properties because the typical 
occupation number is zero or much smaller than unity near the QCP even in 
the soft-core case,
while the difference between the hard and soft constraints manifests 
itself only for the states of which $n_i \ge 2$.

\deffig{Density}{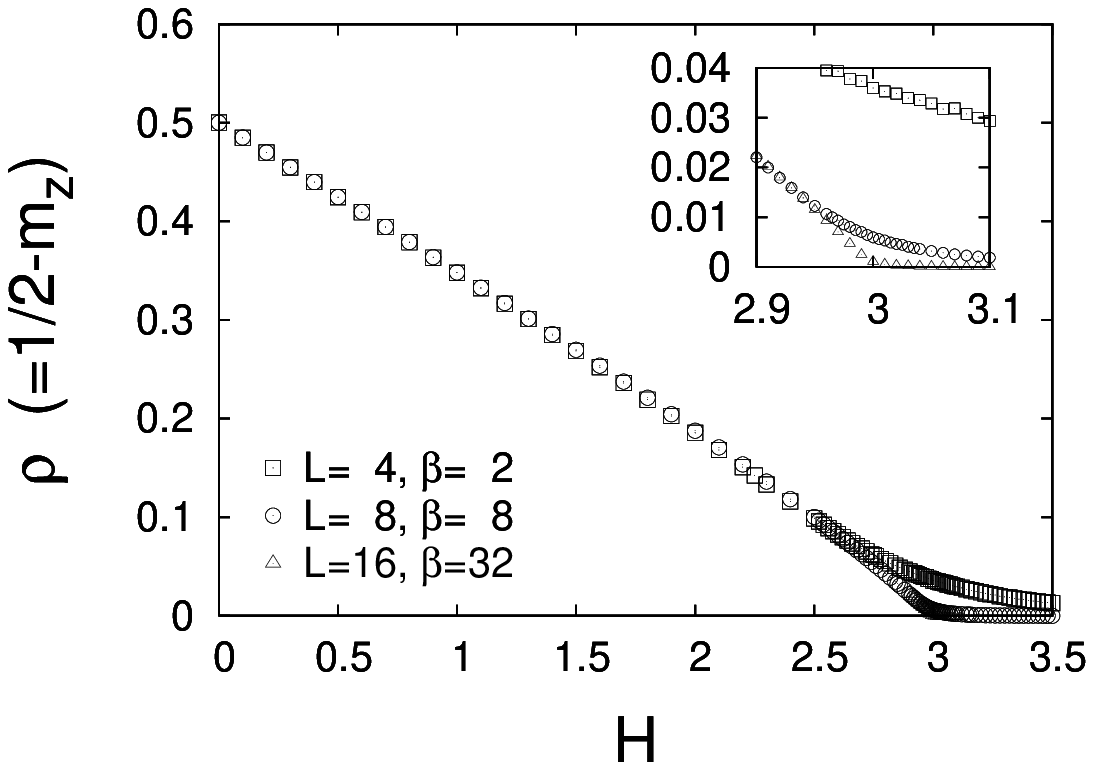}{0.45}{
  Field dependence of magnetization $\rho \equiv \frac12 - \langle S_z \rangle$.
  The three curves correspond to different system sizes, while the
  `scaled' temperature $L^2/\beta$ is fixed to be 8.
  The inset is the close-up view of the quantum critical region.
}

In the following few paragraphs, we present the results obtained 
earlier\cite{Uzunov1981,FisherWGF1989} for the Bose gas
that can be found also in textbooks.\cite{Text}
We reproduce them here since it is convenient to have
explicit formulas in the form directly comparable 
with the present Monte Carlo results.
(In particular, explicit formulas for the system-size 
dependence are hardly found in the literature.)
First, the continuous field theory for model
\Eq{BosonicHamiltonian} is characterized by the action
\begin{eqnarray}
  S & = & \int d\vect{x} \int d\tau\ 
         \left( \rule{0mm}{5mm}
             \psi^*\frac{\partial \psi}{\partial \tau}
           + \left|\nabla\psi\right|^2
         \right. \nonumber \\
    &   &
         \left. \rule{0mm}{5mm}
         \qquad
         - h\,{\rm Re}\,\psi 
         - r \left|\psi\right|^2
         + u \left|\psi\right|^4 
         \right)
         .
         \label{eq:Action}
\end{eqnarray}
The parameter $r$ depends on $\mu$ as $r \propto \mu - \mu_c$
near the QCP, and the symmetry-breaking field $h$, 
which does not correspond to the original magnetic field, 
is included for a technical purpose.
From a simple dimensional analysis, the effect of the scale
transformation up to the scale $b$ turns out to be the following:
\begin{eqnarray}
  & &
      \beta \rightarrow \tilde \beta \equiv  \beta b^{-2}, \ 
      L \rightarrow \tilde L \equiv L b^{-1}, \
      \psi \rightarrow \tilde \psi \equiv \psi b^{\frac{d}{2}}, 
      \nonumber \\ 
  & &
      h \rightarrow \tilde h \equiv h b^{2+\frac{d}{2}}, \
      r \rightarrow \tilde r \equiv r b^2,\ 
      u \rightarrow \tilde u \equiv u b^{2-d}, \nonumber
\end{eqnarray}
where $L$ is the system size.
Hence, the upper critical dimension $d_c$ is equal to 2.\cite{Uzunov1981}
For three dimensions, therefore, we should expect a
mean-field-type scaling.

\deffig{Susceptibility}{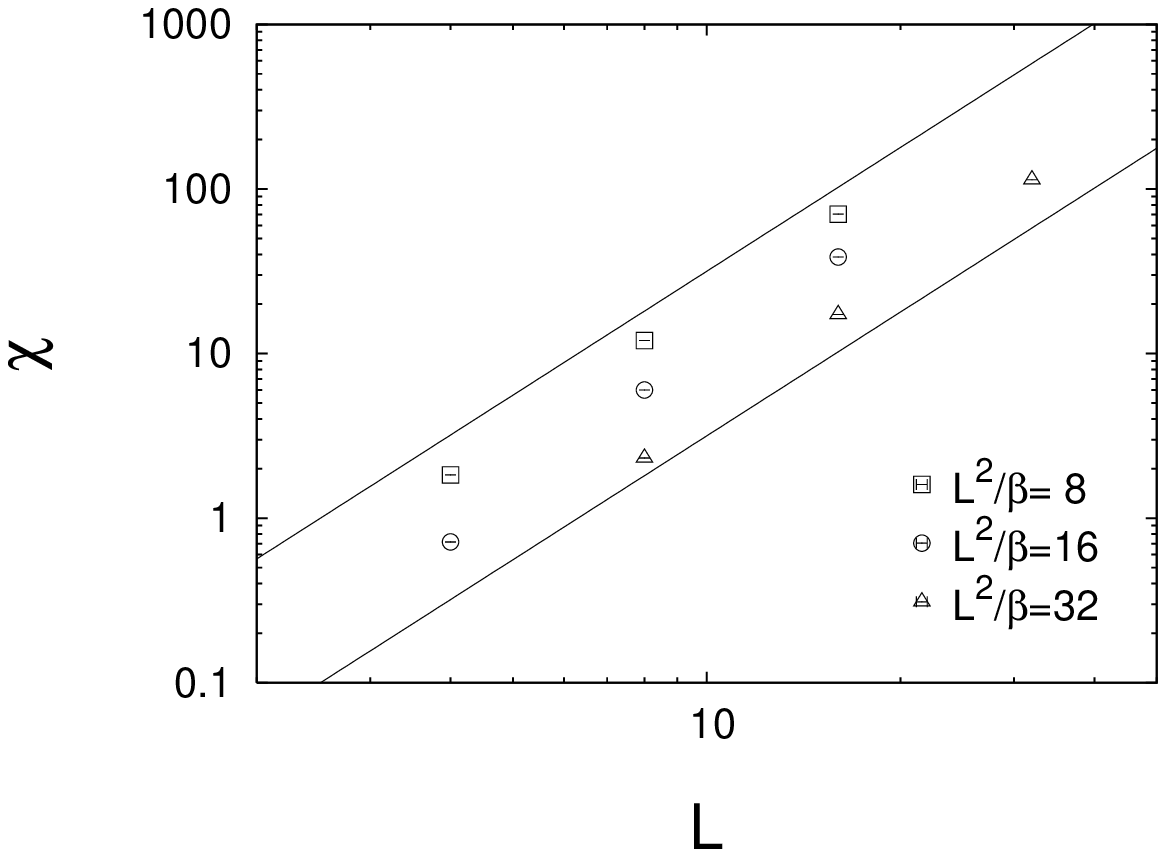}{0.40}{
  Size dependence of susceptibility 
  at quantum critical point, $H=3J$.
  Three values of the scaled inverse temperature, 
  $\beta/L^2=8,16$ and $32$, 
  are examined. The solid lines indicate the slope
  of $\chi \propto L^{2.5}$.
}

Because of the dangerous irrelevant parameter $u$,
the usual procedure of the finite size scaling, i.e.,
rescaling the coordinate and the abscissa by size-dependent 
factors, does not work.
We first define $\Phi(h,r,u,\beta,L)$ as the singular 
part of $-\log Z(h,r,u,\beta,L)$.
We assume that $\Phi$ has the property
$
  \Phi(h,r,u,\beta,L) 
  =
  \Phi(\tilde h, \tilde r, \tilde u, \tilde \beta, \tilde L)
$
for an arbitrary $b$.
With this expression, we first consider the particle density $\rho$,
which corresponds, in the original model, 
to the deviation of the magnetization (parallel 
to the magnetic field) from its saturation value.
The particle density is related to the magnetization
perpendicular to the field, which we denote as $m$,
by $\rho \sim m^2 \sim |\psi|^2$.
Then we have
\begin{eqnarray*}
  m(r,u)
  & \equiv &
    \lim_{h\to 0} \lim_{L\to \infty} \lim_{\beta\to\infty}
    L^{-d} \beta^{-1} 
    \frac{\partial\Phi}{\partial h} \\
  & = &
    b^{-\frac{d}{2}} \tilde m(\tilde r, \tilde u),
\end{eqnarray*}
where $\tilde m(\tilde r, \tilde u)$ is the magnetization of the
renormalized system.
When the system is renormalized up to the point where the correlation 
length is $O(1)$, 
we can neglect fluctuations and apply the mean-field-type calculation
without a serious error in estimating $\tilde m$, which yields
$$
  \tilde m(\tilde r, \tilde u) \sim \sqrt{\frac{\tilde r}{\tilde u}} 
  = \sqrt{\frac{r}{u}} \ b^{\frac{d}{2}}
$$
for sufficiently large $b$. Thus, we arrive at\cite{Uzunov1981}
\begin{equation}
  \rho \propto r \propto |H-H_c| \quad (T=0).
  \label{eq:Linear}
\end{equation}

Similarly, we can obtain the scaling form of 
the susceptibility (of the spin components parpendicular to
the magnetic field) at $r=0$:
\begin{eqnarray}
  \chi(u,\beta,L)
  & \equiv & 
    \lim_{h\to 0} \lim_{r\to 0}  
    \int d\vect{x} \int_0^{\beta} d\tau \ 
    \langle \psi^*(x,\tau) \psi(0,0) \rangle
    \nonumber \\
  & \propto &
    b^2 \tilde \chi(\tilde u, \tilde \beta, \tilde L).
    \nonumber
\end{eqnarray}
For sufficiently large $b$, we have
$$
  \tilde \chi(\tilde u, \tilde \beta, \tilde L) 
  \sim \sqrt{\frac{\tilde\beta}{\tilde u}}
  \sim \sqrt{\frac{\beta}{u}}\ b^{\frac{d}{2}-1},
$$
where $L$ is assumed to be larger than $b$.
Therefore, the asymptotic system size dependence of
the critical susceptibility for a fixed value of
the scaled inverse temperature $\beta' \equiv \beta / L^2$ is
\begin{equation}
  \chi(u, \beta' L^2, L)
  \propto L^{1+\frac{d}{2}}.
  \label{eq:Susceptibility}
\end{equation}

Finally, we consider the temperature dependence of the
critical magnetic field.
When $d>2$, we can make $\tilde u$ small by choosing large $b$
so that the perturbation theory in $\tilde u$ is nearly exact 
for the renormalized system.
The perturbation theory indicates that $\Phi$ for the
renormalized system has a 
singularity when the condition\cite{Text}
$
  \tilde r = \tilde r_c \equiv A \tilde u / \tilde \beta^{d/2}
$
is satisfied with $A$ being a numerical constant.
This means that in terms of the bare coupling constants,
the singularity occurs when\cite{Uzunov1981}
$$
  H_c(T)-H_c(0) \propto T^{d/2}.
$$

\deffig{FSS}{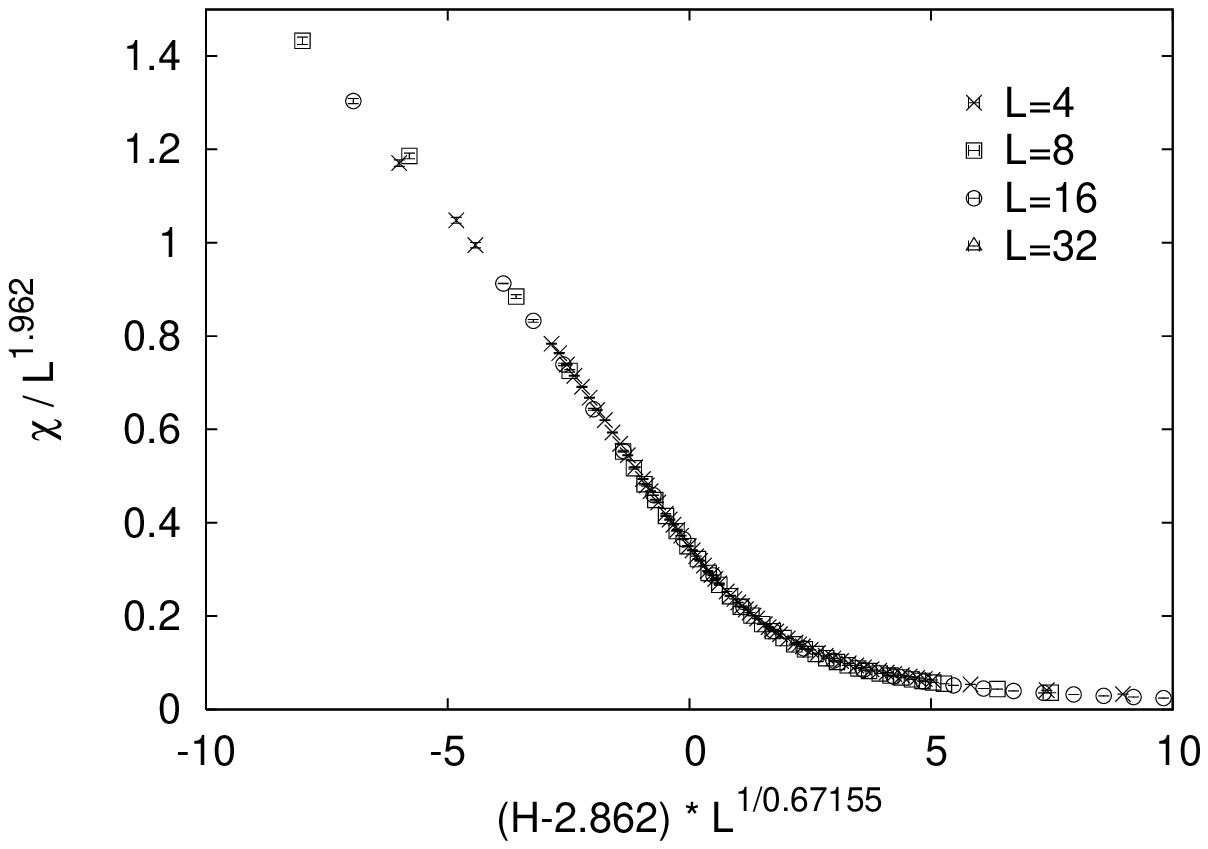}{0.40}{
  Finite size scaling plot of susceptibility
  at $\beta=4J$.
  The critical exponents used here are taken from
  the results of the classical $XY$ model.
  \cite{CampostriniHPRV2001}
}

\deffig{PhaseDiagram}{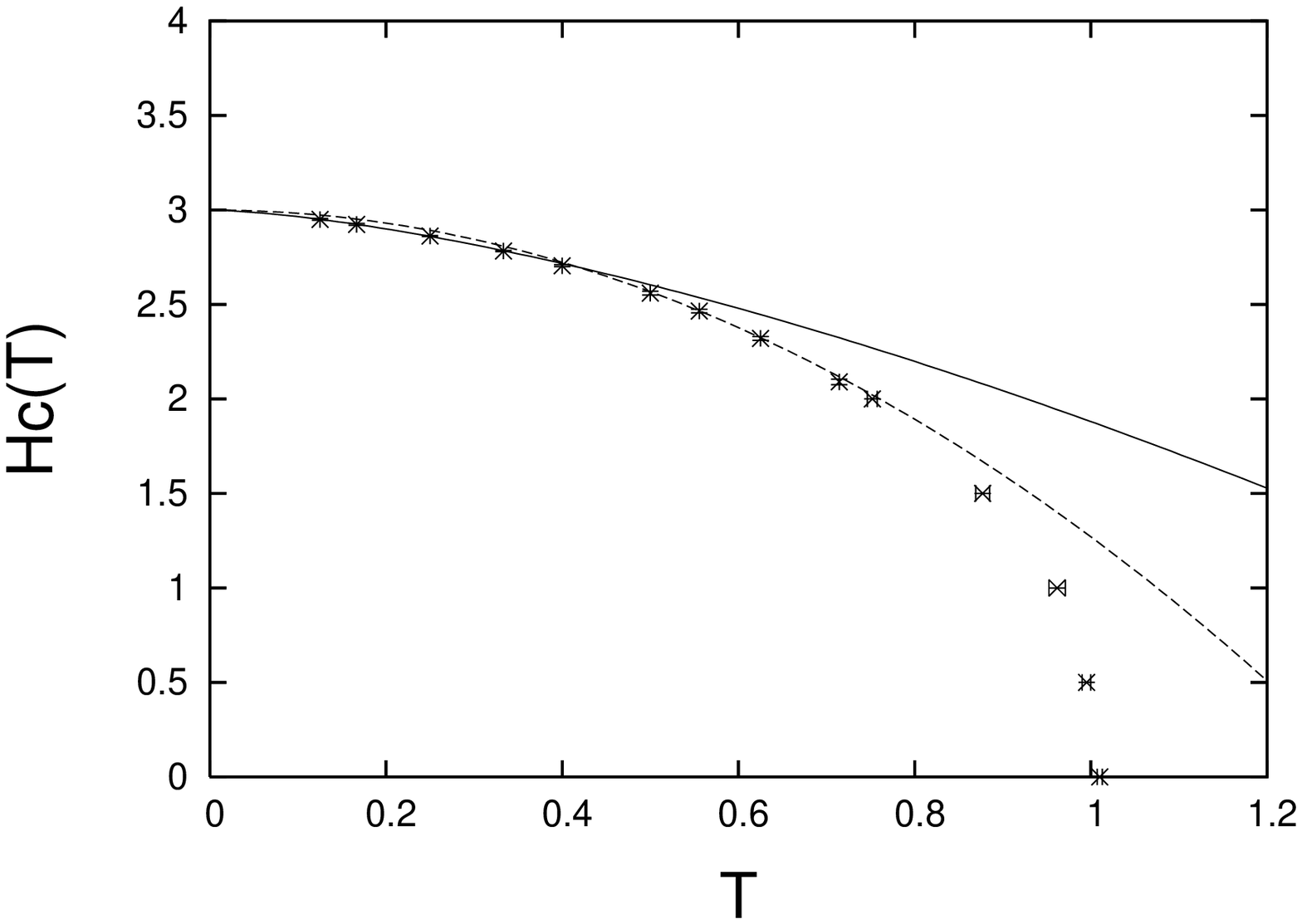}{0.40}{
  Critical field as a function of temperature.
  The solid curve corresponds to $\phi= 1.5$, whereas
  the dashed curve $\phi=2.0$.
}

We now present the results of our Monte Carlo simulation
for spin Hamiltonian \Eq{XYHamiltonian} on a simple cubic lattice
and compare them with the theoretical predictions.
The simulation method employed here is based on
the directed-loop algorithm
\bcite{SyljuasenS} for which a review can be found in
a recent article.\bcite{KawashimaH2004}
The system sizes that we explored range from $L=4$ to $L=32$ and the
temperatures from $\beta=0.5$ to $\beta=32$ (for $L=32$).

\Figure{Density} shows the magnetic field dependence of the deviation of
the longitudinal magnetization from its saturation value.
In the bosonic language, this is the particle density which is a function
of chemical potential. 
Results are shown only for three system sizes $L=4,8$ and $16$
with fixed `scaled' temperature, $L^2/\beta=8$.
As can be seen in the figure, the system size dependence remains 
only in the vicinity of the quantum critical point.
The inset shows the magnified view of the critical region.
The linear field dependence \Eq{Linear} can be clearly observed
in this inset.

Next, we show the susceptibility of the perpendicular 
components of spins. This is defined as 
$\chi \equiv \sum_{\vect r}\int_0^{\beta} d\tau
\langle S^x(\vect r, \tau) S^x(\vect 0, 0) \rangle$
for model \Eq{XYHamiltonian}.
In \Fig{Susceptibility}, the system size dependence of the
susceptibility at the critical point $H=3J$ is shown.
Three different values of the scaled temperature $L^2/\beta$
are chosen.
For each value of the scaled temperature,
the asymptotic size dependence is well fitted by the straight 
line with the slope $2.5$, in agreement with \Eq{Susceptibility}.
Thus the universality class of the QCP is identified as
the mean field type.

In order to obtain the phase diagram in the $H-T$ plane,
we carry out the finite-size scaling of the susceptibility
with a fixed magnetic field or a fixed temperature.
As an example, a finite-size-scaling plot is
shown for $\beta J=4$ in \Fig{FSS}.
The data can be scaled nicely with the critical exponents
estimated for the classical $XY$ model, namely,
$\nu = 0.67155(27)$ and $\eta = 0.0380(4)$.\cite{CampostriniHPRV2001}
The value of the critical field is the only
free parameter determined by the present data,
and the best plot is obtained for
$H_c(T=4J) = 2.862$.
In this manner, we estimated the critical field (temperature) at 
various temperatures (fields).

\deffig{CriticalField}{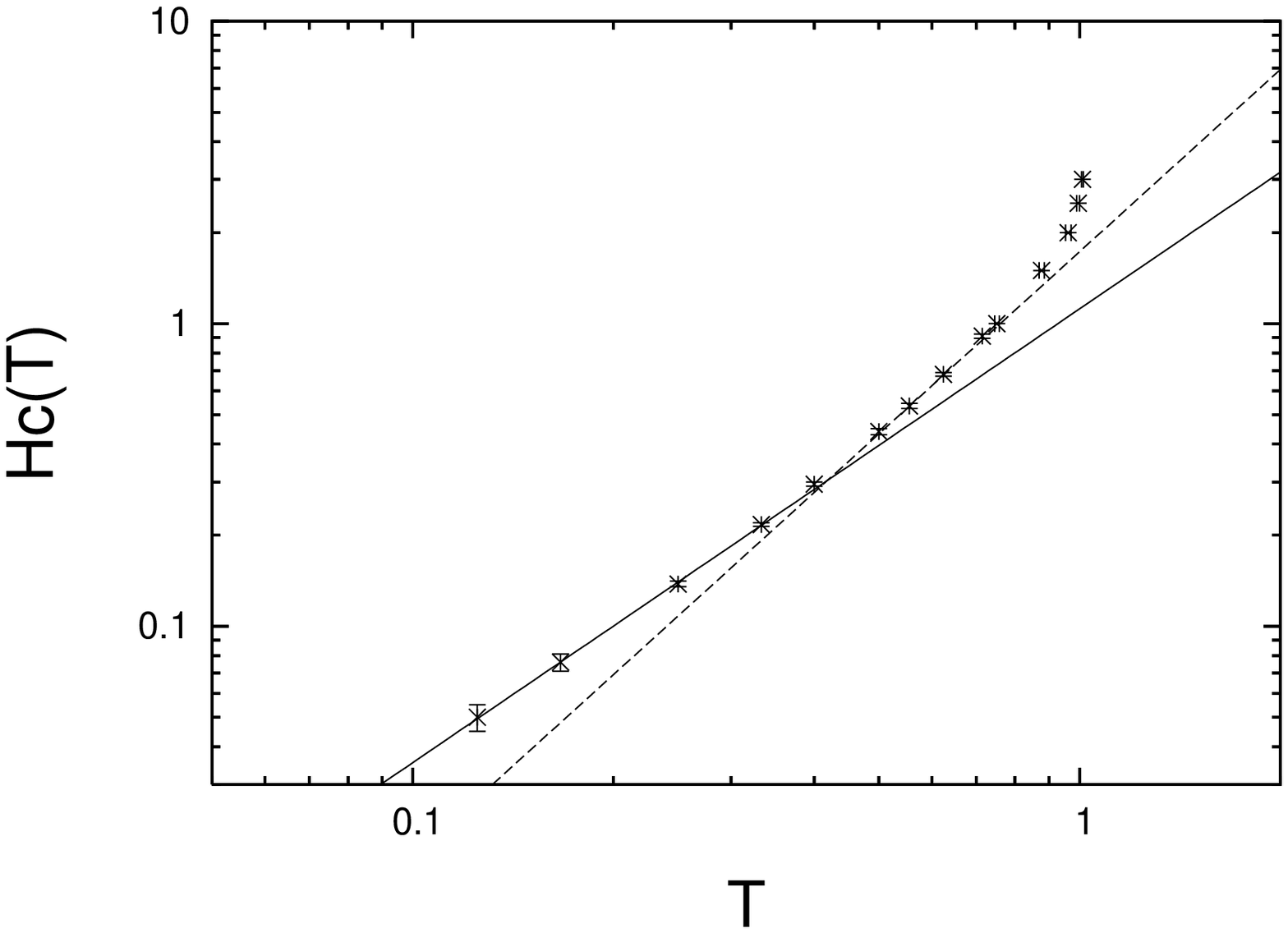}{0.40}{
  Logarithmic plot of critical field against temperature.
  The two straight lines correspond to $\phi = 1.5$ (solid)
  and $\phi = 2.0$ (dashed).
  In the intermediate temperature region, $\phi=2.0$ seems
  to fit the data well, while in the low-temperature region, 
  the correct slope $\phi = 1.5$ yields a better fitting.
}

In \Fig{PhaseDiagram}, two curves are plotted for comparison.
The solid curve represents the mean-field critical exponent,
$\phi = 1.5$, whereas the dashed curve represents the previous estimate
\cite{TlCuCl2001} based on an experiment on TlCuCl$_3$,
$\phi = 2.0$.
At a first glance, it seems that the curve with $\phi=2.0$
fits the data better than that with $\phi=1.5$.
However, when the logarithmic scale is used, as we do in
\Fig{CriticalField}, it is rather clear that $\phi=2.0$
explains only a transient behavior, and the correct
asymptotic value of the exponent is $\phi=1.5$.

To summarize, we have argued that
the mean-field exponents should be correct and exact for the
quantum critical point of the $XY$ model in three dimensions,
and our numerical simulations demonstrated 
that it is indeed the case beyond reasonable doubt.
Therefore, we can at least exclude 
the inaccuracy of the Hartree-Fock approximation 
as the source of the discrepancy between the theoretical
and experimental results.
We have also found that there is a relatively large 
intermediate temperature region where a transient behavior
can be well described by the effective exponent $\phi\sim 2$.
While there could be some other possibilities that may truly 
change the critical behavior from the mean field type to something else,
such as a presence of Dzyaloshinsky-Moria interactions, 
the present results suggest that it is possible to
explain the discrepancy between the theoretical and experimental results
only by taking into account the transient behavior
and that the quantum critical behavior
driven by the magnetic field can be correctly described 
by the $XY$ model (or the diluted Bose gas fixed point)
which yields the mean-field exponents.


The author thanks M.~Oshikawa and H.~Otsuka for helpful comments.
This work was supported by a grant-in-aid (Program No.\ 14540361)
from Monka-sho, Japan. 
The computation presented here was performed using the computers at
Supercomputer Center, Institute of Solid State Physics, University of Tokyo.


\end{document}